# Thermal Insulating Polymer-Air Multilayer for Window Energy Efficiency

Rui Kou,[1,†] Ying Zhong,[1,2,†] Qingyang Wang,[3] Jeongmin Kim,[3] Renkun Chen,[3,4] Yu Qiao[1,4,*]

[1] *Department of Structural Engineering, University of California – San Diego, La Jolla, CA 92093-0085, United States*

[2] *Department of Mechanical Engineering, University of South Florida, Tampa, Florida 33620, United States*

[3] *Department of Mechanical and Aerospace Engineering, University of California – San Diego, La Jolla, CA 92093, United States*

[4] *Program of Materials Science and Engineering, University of California – San Diego, La Jolla, CA 92093, United States*

[*] *Corresponding author. Email: yqiao@ucsd.edu.* [†] Those authors contribute equally.

## Abstract

Polymer-air multilayer (PAM) was developed to decrease the heat loss through window glass panes. A PAM consists of a few polymer films separated from each other by air gaps. Thanks to the excellent optical properties of the polymer films, the visual transmittance of PAM is higher than 70%, and the haze is less than 2%. PAM not only has mechanisms to reduce the conductive and convective heat transfer, but also can obstruct the radiative heat transfer. With a 4~6 mm thick PAM coating, the *U*-factor of a glass pane can be lowered from above 1 Btu/(h· $ft^2$ · °F) to 0.5~0.6 Btu/(h· $ft^2$ · °F). PAM is resilient and robust, relevant to the window retrofitting applications.

## Nomenclature

| | |
|---|---|
| $A$ | Area [$m^2$] |
| $CRI$ | Color rendering index |
| $d$ | Distance [m] |
| $\Delta E_{\mathrm{uvw}}$ | Color difference |
| $E(\lambda)$ | Solar spectral irradiance [$W/m^2$] |
| $h$ | Heat transfer coefficient [$W/(m^2K)$] |
| $H$ | Sample height [m] |
| $Nu$ | Mean Nusselt number |
| $q$ | Heat flux per unit area [$W/m^2$] |
| $R$ | Thermal resistance [$m^2K/W$] |
| $T$ | Temperature [K] |
| $t$ | Thickness [m] |
| $U$ | Overall thermal transmittance [$Btu/(h \cdot ft^2 \cdot °F)$] |
| $V_{\mathrm{T}}$ | Visual transmittance |
| $x$, and $y$ | CIE 1931 chromaticity |
| $X$, $Y$ and $Z$ | CIE tristimulus values |
| $\varepsilon$ | Surface emissivity |
| $\kappa$ | Thermal conductivity [$W/(m^2K)$] |
| $\lambda$ | Wavelength [nm] |
| $\tau$ | Transmittance |

*Sub-indices*

| | |
|---|---|
| cv | Convection |
| cd | Conduction |
| ex | Exterior |
| G | Glass |
| gap | Gap between films |
| in | Interior |
| r | Radiation |
| ST | Standard |

## 1. Introduction

Enhancing the building energy efficiency, particularly minimizing the heat loss through building windows, is of immense importance. Among all the consumed energy related to buildings, 40% is used for building heating, ventilation, and air conditioning (HVAC), in which ~25% is lost because of the inefficient thermal insulation of windows [1]. About 30~40% of present-day windows in the U.S. are still single-paned and they cause nearly 50% of the total window thermal energy loss [2]. In the next 40 years, hardly could they be fully replaced by more energy efficient insulated glass units (IGU) [2][3]. It is envisioned that, if highly transparent, low-haze, and highly thermal insulating layers can be produced and attached to single-pane windows, the energy saving and the improvement in occupant comfort would be significant.

Low-cost, transparent, and highly thermal insulating coating materials are still lacking. Silica aerogel, for example, can have a low thermal conductivity ($\kappa$) of ~0.01 W/(m·K) [4,5], partly due to the high porosity (> 99%) and partly due to the Knudsen effect of the rarified air in the nanopores [6,7]. Its visual transmittance can be satisfactory, as the nanopore size is much smaller than the visual light wavelength [8,9]. However, with the structural integrity being considered, the porosity must be relatively low and therefore, the ultralow thermal conductivity may not be realized. With a porosity ~50%, the thermal conductivity would be 0.04~0.1 W/(m·K), higher than that of air [8]. Moreover, haze of silica aerogel tends to be large, due to the pore size distribution [6,7,10–12]. Low-emissivity (low-e) film, for another example, blocks radiative heat transfer in the infrared (IR) range. An issue of low-e film is water condensation. Low-e film does not reduce conductive heat transfer. If its surface is in contact with a water layer, heat conduction would dominate the thermal properties [13–15]. The large decrease in visual transmittance, the change in color [16], and the relatively high materials and installation costs [17] are also of major concerns.

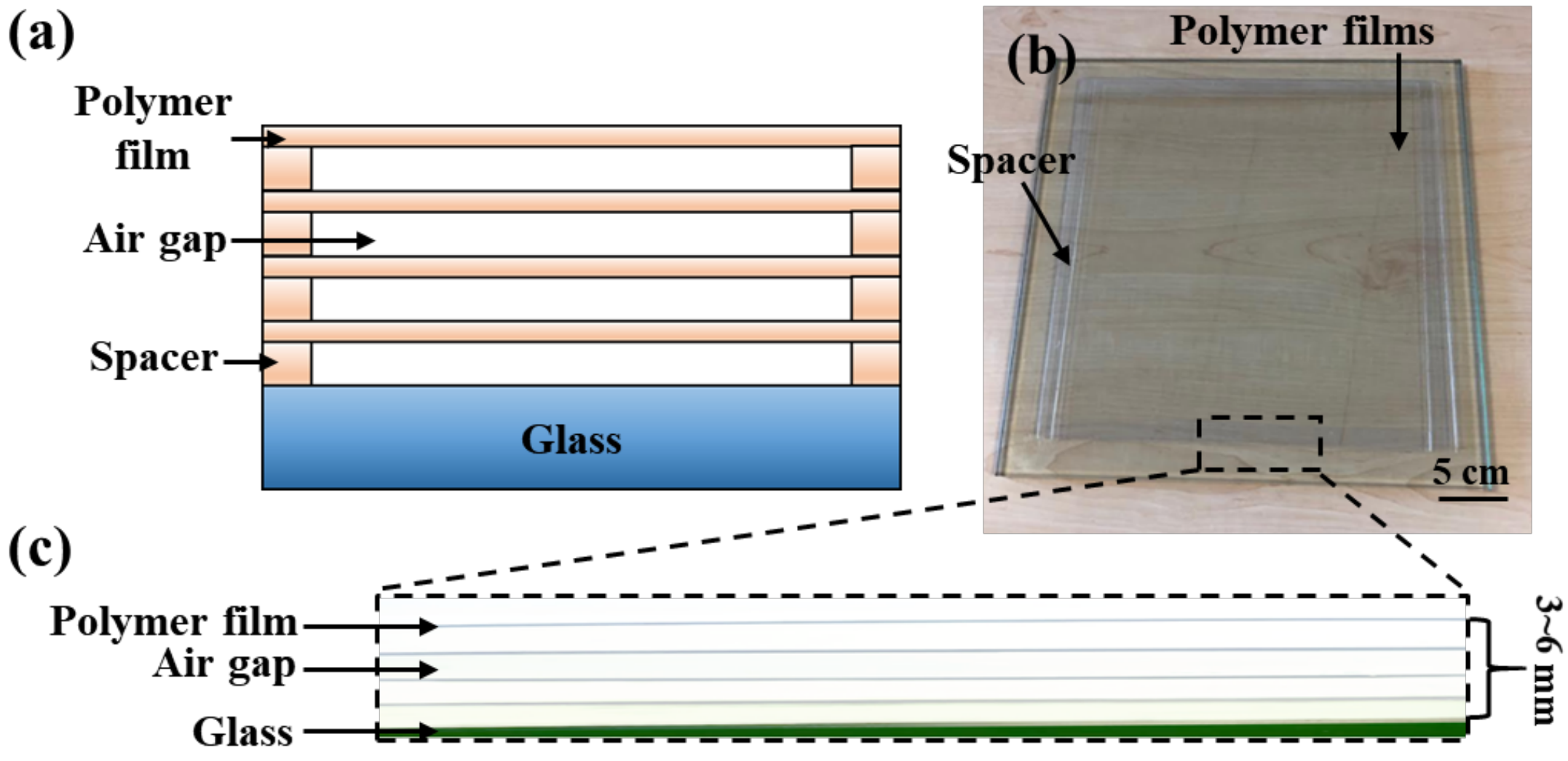

**Figure 1** (a) Schematic of the polymer-air multilayer (PAM). Photos of (b) top view and c) side view of a typical PAM sample.

In the current research, we develop and characterize a polymer-based coating structure, referred to as the polymer-air multilayer (PAM). Figure 1(a) depicts a PAM, which comprises multiple transparent polymer films separated by air gaps. The polymer layer thickness may range from less than 10 μm to more than 100 μm, and the air gap thickness may range from less than 0.1 mm to more than 1 mm. The layer separation is achieved by slightly stretching the polymer films, aided by surface electrification [18,19]. As the air gaps are thinner than 1 mm, convection heat transfer of air is negligible [20]. The air gaps have a low thermal conductivity ~0.026 W/m·K. In addition, each layer of polymer can block a portion of the IR radiation heat transfer due to the absorption and bi-directional re-emission of the radiant flux in each layer. Because all the conductive, radiative, and convective heat transport rates are reduced, we envision that PAM would possess excellent thermal insulation properties. Its optical properties, such as visual transmittance and haze, are dominated by the transparency and the flatness of the polymer films.

## 2. Experimental Procedure

To prove the concept, we produced PAM samples by using 4 layers of 125-μm-thick 0.3-m-large electrified polyethylene terephthalate (PET) films. The untreated PET films were obtained from McMaster-Carr (Product No. 8567K52). The spacers were made of 6.35-mm-wide polycarbonate (PC) bars (McMaster-Carr, Product No. 85585K15). The spacers were placed in between adjacent PET films at the edges. The layer stack was mounted on a 10-mm-thick 0.37-m-large square glass pane, as shown in Fig.1 (b,c). The spacer thickness ranged from 500 μm to 1 mm, which determined the thickness of the air gaps. The spacers were affixed by adhesives (3M CA5) that cured at ambient temperature. The PAM assembly and mounting procedure will be discussed in the Installation section below.

### 2.1 Polymer electrification

The PAM samples were formed by surface-electrified PET films. The electrification was conducted through corona charging and liquid electrification, following the work of Zhong et al. [18,19]. In the first step, isopropyl alcohol (IPA) was used to ultrasonically clean the as-received PET films for about 15 min. Then, they were cleaned in de-ionized (DI) water for another 5 min and dried in vacuum at 60 ℃ for 24 hr. A 19-mm-long 1.5-mm-diameter tungsten needle electrode and a steel-wire grid mesh were connected to two Glassman FJ Series 120 Watt regulated high-voltage DC power supplies, with the capacities respectively being 40 kV and 20 kV [18]. A PET film was firmly placed and flattened on a grounded stainless-steel plate. The grid mesh was 4 mm away from the polymer surface. The distance between the needle tip and the grid was 4 cm. The voltage on the needle electrode was -10 kV; the charging time was 60 sec. After the charged PET film was lifted from the grounded steel plate, the positive side was covered by a polycarbonate container. The negative side was dipped into an aqueous solution of sodium formate (SF) and dried in air. The SF concentration was 10 mM. All the experiments were carried out in lab air at room temperature, with the relative humidity ~60%. Surface voltages of the PET film were monitored by a Trek Model 344 voltmeter from both sides, with the probe distance of 25 mm. The measurement result suggested that the dipolar charge density was around 0.5 mC/m$^2$.

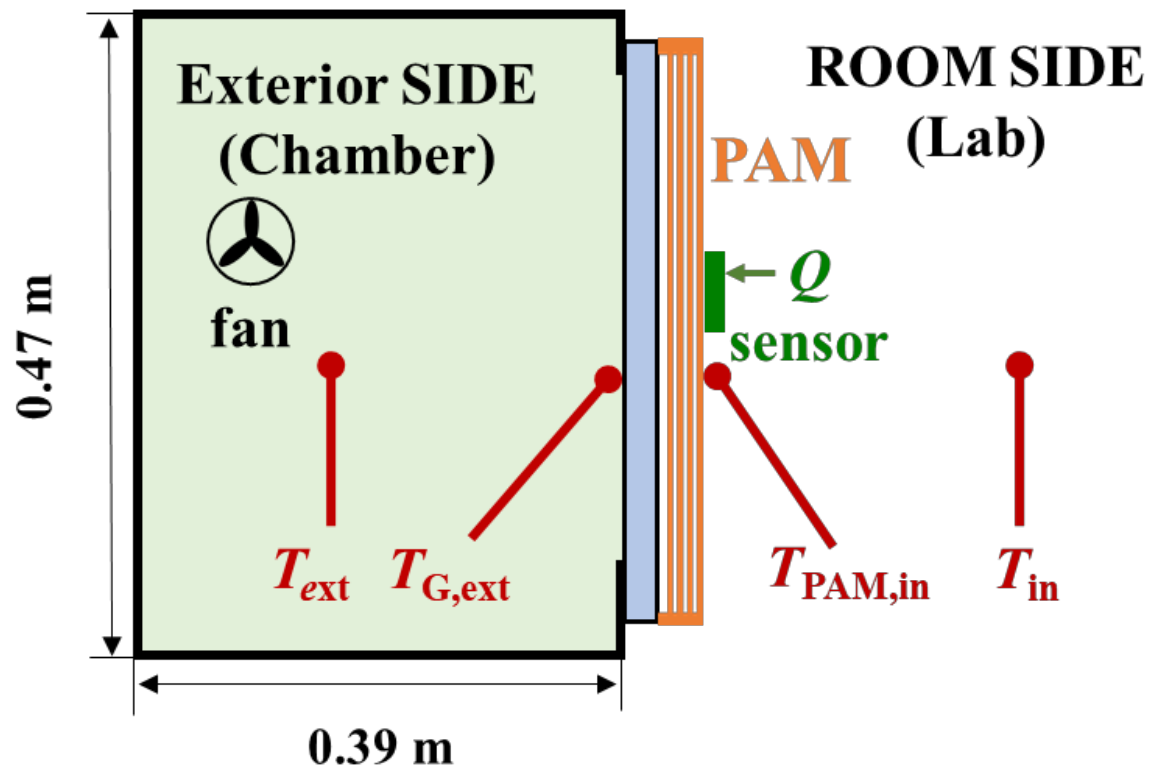

**Figure 2** Schematic of the *U*-factor measurement setup.

2.2 Thermal transmittance (*U*-factor) measurement

The rate of heat loss through windows is commonly characterized by the *U*-factor. As shown in Figure 2, a 0.47-m-tall 0.39-m-long 0.47-m-wide environmental chamber was employed for the *U*-factor measurement, consistent with ASTM C1199, ASTM C518, NFRC-102, and NFRC-102. Inside the chamber, the temperature ($T_{\text{ex}}$, simulating the exterior temperature of the weather side of a window) was kept at about -18 °C. Outside the chamber, the ambient temperature ($T_{\text{in}}$, simulating the interior room side temperature) was ~21°C. A PAM sample was mounted on the 0.37×0.37 m$^2$ large glass window in the chamber wall. To monitor the heat flux, a gSKIN XI heat flux sensor was affixed on the room side surface of the PAM sample. $T_{\text{ex}}$ and $T_{\text{in}}$, as well as the temperatures at the innermost surface of PAM and the outermost surface of glass ($T_{\text{PAM,in}}$ and $T_{\text{G,ex}}$) were monitored by four Omega type-K thermocouples, respectively. Based on ASTM C1199 and NFRC-102, for the purpose of calculating the standard *U*-factor, the standard heat resistance to the exterior side was $R_{\mathrm{ST,ex}} = {h_{\mathrm{ST,ex}}}^{-1} = 1/30$ m²·K/W and the interior side thermal resistance ($R_{\mathrm{ST,in}}$) is

$$R_{\mathrm{ST,in}} = \frac{1}{h_{\mathrm{ST,in}}} = \left\{1.46\left[\frac{\left(T_{\mathrm{in}} - T_{\mathrm{PAM,in}}\right)}{H}\right]^{0.25} + \sigma_{\mathrm{SB}}\varepsilon\left[\frac{(T_{\mathrm{in}} + 273.16)^4 - \left(T_{\mathrm{PAM,in}} + 273.16\right)^4}{\left(T_{\mathrm{in}} - T_{\mathrm{PAM,in}}\right)}\right]\right\}^{-1} \tag{1}$$

where $H$ represents the height of the PAM, $\sigma_{\mathrm{SB}} = 5.67 \times 10^{-8}\ \mathrm{W/m^2 \cdot K^4}$ is the Stefan Boltzmann constant, and $\varepsilon$ represents the surface emissivity of the innermost surface of polymer film [15]. The thermal transmittance ($U$-factor) is equal to the reciprocal of the total thermal resistance,

$$U = \left(R_{\mathrm{ST,ex}} + \frac{T_{\mathrm{PAM,in}} - T_{\mathrm{PAM,ex}}}{q} + R_{\mathrm{ST,in}}\right)^{-1} \tag{2}$$

2.3 Measurement of optical properties

With a JASCO V770 UV-VIS spectrometer, visual transmittance ($V_{\mathrm{T}}$) measurement was carried out for the spectral range from 380 nm to 780 nm, according to ASTM D1003. The PAM sample was located in front of an integrating sphere, in which the visible light could be collected by a photo detector. The UV-Vis spectrophotometer was utilized to measure the visual light intensity passing through the PAM sample and compare it to the incident light intensity. The visible light transmittance ($V_{\mathrm{T}}$) is defined in ASTM G173 as

$$V_T = \frac{\int \tau(\lambda)E(\lambda)d\lambda}{\int E(\lambda)d\lambda}, \tag{3}$$

among which $E(\lambda)$ is the solar spectral irradiance, $\tau(\lambda) = \tau_t/\tau_i$, $\tau_t$ indicates the light passed through the PAM sample, and $\tau_i$ represents the incident light. Haze was calculated from the sample diffusion $\tau_{s,d}$ and the instrument diffusion $\tau_{i,d}$:

$$\text{Haze}= \tau_{s,d}/\tau_t - \tau_{i,d}/\tau_{i.} \quad (4)$$

The transparency color perceptions of PAM was depicted by the CIE 1931 chromaticity diagram, designed to represent the mapping of human color perception. The CIE tristimulus values characterize the light intensity according to three primary colors. They were obtained from a CIE 1931 standard observer [21][22], typically represented by the *X*, *Y* and *Z* (ASTM E308):

$$X = \int R(\lambda)SPD(\lambda)\bar{x}(\lambda)d\lambda \quad (5)$$

$$Y = \int R(\lambda)SPD(\lambda)\bar{y}(\lambda)d\lambda, \quad (6)$$

$$Z = \int R(\lambda)SPD(\lambda)\bar{z}(\lambda)d\lambda. \quad (7)$$

where $\bar{x}(\lambda)$, $\bar{y}(\lambda)$, and $\bar{z}(\lambda)$ are the color matching functions of the chromatic response of the standard observer, $R(\lambda)$ is the reflectance, transmittance or radiance factor, and *SPD*($\lambda$) represents relative spectral power distribution (SPD) [22]. The CIE 1931 (x, y) chromaticity values are

$$x = \frac{X}{X+Y+Z} \quad (8)$$

$$y = \frac{Y}{X+Y+Z}. \quad (9)$$

The CIE 1931 (x, y) chromaticity value under illumination can be converted into CIE 1964 color space by:

$$u = \frac{4X}{X+15Y+3Z} \quad (10)$$

$$v = \frac{6Y}{X+15Y+3Z}. \quad (11)$$

The CIE 1965 (*U**, *V**, *W**) can be calculated by (CIE 1960 UCS [23]):

$$W^* = 25Y^{1/3} - 17$$

$$U^* = 13W^*(u - u_0) \tag{12}$$

$$V^* = 13W^*(v - v_0), \tag{13}$$

in which $(u_0,\ v_0)$ is the white point. The color difference can be obtained by:

$$\Delta E_{\mathrm{uvw}} = \sqrt{(U^*)^2 + (V^*)^2 + (W^*)^2}, \tag{14}$$

The color rendering index (*CRI*) is [22,23]

$$CRI = 100 - 4.6\Delta E_{\mathrm{uvw}}. \tag{15}$$

2.4 Structural properties of PAM

Robustness and resilience tests were performed on the 4-layer PAM samples with the initial thickness of ~4 mm. The air gap thickness was ~0.875 mm. Compression test was conducted by using a Type 5582 Instron machine. The loading rod had a 5-mm-radius round tip. The crosshead was controlled in the displacement mode, with the speed of 10 μm/s. The total compression displacement was 3.5 mm. 4000 loading-unloading cycles were applied and the thickness at the center point was measured periodically by a micrometer. To monitor the long-term stability, a PAM sample was placed in open lab air for one year (relative humidity: 60%), and the center-point thickness was frequently measured.

## 3. Modelling of the *U*-factor

Heat transfer through PAM depends on the heat conduction ($R_{\mathrm{cd}}$), air convection in air gaps and at exterior surface ($R_{\mathrm{cv}}$), and the radiative heat flow across the reflective polymer film surfaces and glass pane ($R_{\mathrm{r}}$), as shown in Figure 3. Because the in-plane PAM size is more than 50 times larger than its thickness, the heat transfer is quasi-one-dimensional [20]. Due to the IR opacity of polymer films in long-wavelength infrared, we assume no transmission of radiation or internal radiation within the film. Assuming that there are *N* layers of polymer films and one glass pane, and each surface of film and glass pane is isothermal, the steady-state thermal balance equation of the nodes is:

$$q_{\mathrm{r,ex}} + q_{\mathrm{cv,ex}} = q_{\mathrm{cd,G}} = q_{\mathrm{r,gap,i}} + q_{\mathrm{cd,gap,i}} + q_{\mathrm{cv,gap,i}} = q_{\mathrm{cd,P,i}} = q_{\mathrm{r,in}} + q_{\mathrm{cv,in}} \tag{16}$$

where $q_{\mathrm{r,ex}}$ and $q_{\mathrm{cv,ex}}$ are radiative and convective heat fluxes from the weather side to6the exterior window surface, respectively; $q_{\mathrm{cd,G}}$ and $q_{\mathrm{cd,P,i}}$ are conductive heat fluxes through the glass and the *i*-th layer of polymer film, respectively; $q_{\mathrm{r,gap,i}}$, $q_{\mathrm{cd,gap,i}}$ and $q_{\mathrm{cv,gap,i}}$ are the radiative, the conductive, and the convective heat fluxes, respectively, through the air gaps; $q_{\mathrm{r,in}}$ and $q_{\mathrm{cv,in}}$ are respectively the radiative and the convective heat fluxes from the innermost polymer film to the room side.

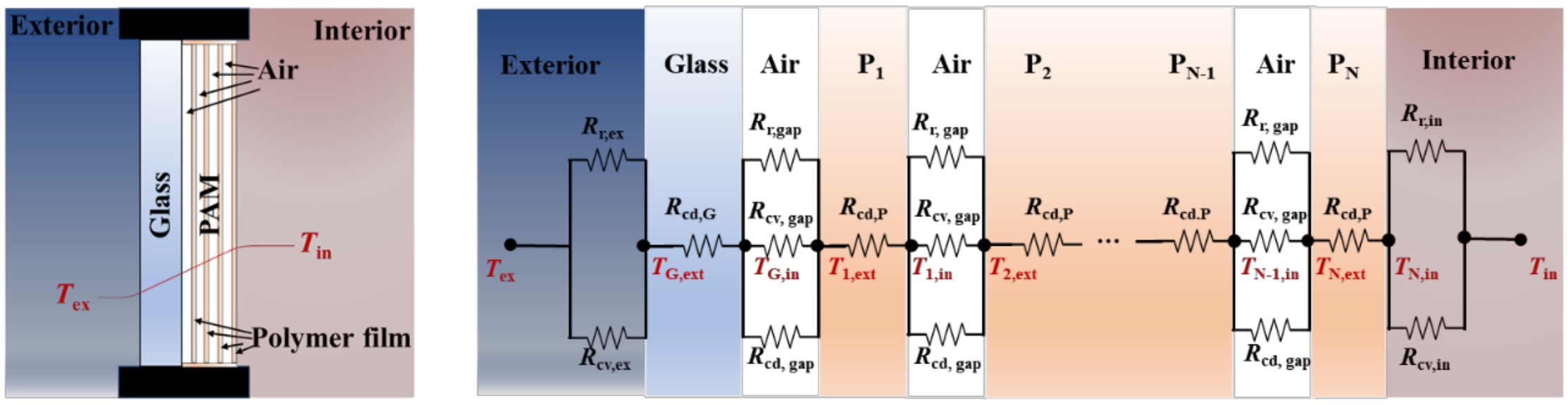

**Figure 3** The model of the thermal resistance network of PAM. The temperature nodes include the following: $T_{\mathrm{ex}}$ (the exterior temperature), $T_{\mathrm{G,ex}}$ and $T_{\mathrm{G,in}}$ (the exterior and the interior temperatures of the windowpane), $T_{\mathrm{i,ex}}$ and $T_{\mathrm{i,in}}$ (the exterior and the interior temperatures of the *i*-th layer of the polymer film), and $T_{\mathrm{in}}$ (the interior temperature). The exterior weather side temperature and the room side temperature are set to $T_{\mathrm{ex}} = -18$ °C and $T_{\mathrm{in}} = 21$°C.

The conductive heat transfer through glass, air, and polymer follows the Fourier law [24],

$$q_{\mathrm{cd}} = -\kappa \cdot \nabla T \tag{17}$$

in which $\nabla T$ is the temperature gradient and $\kappa$ is the thermal conductivity of the respective medium. The radiative heat transfer between two polymer layers is governed by the Kirchhoff Law [20],

$$q_{\mathrm{r},i} = \frac{\sigma_{\mathrm{SB}}\left({T_{i,\mathrm{ex}}}^4 - {T_{i-1,\mathrm{in}}}^4\right)}{\frac{1}{\varepsilon_{i,\mathrm{ex}}} + \frac{1}{\varepsilon_{i-1,\mathrm{in}}} - 1} \tag{18}$$

where the Stefan Boltzmann constant $\sigma_{\mathrm{SB}} = 5.67 \times 10^{-8}\,\mathrm{W/(m^2K^4)}$. The radiative heat transfer rate on the interior surface of the innermost layer and the exterior surface of window glass can be calculated by:

$$q_{\mathrm{r,in}} = \varepsilon_{\mathrm{P}}\sigma\left({T_{\mathrm{in}}}^4 - {T_{N,\mathrm{in}}}^4\right) \tag{19}$$

$$q_{\mathrm{r,ex}} = \varepsilon_{G}\sigma\left({T_{G,\mathrm{ex}}}^4 - {T_{\mathrm{ex}}}^4\right)$$

For the convective heat transfer in air gaps and at the outermost surfaces of window glass and polymer film, the heat flux is

$$q_{\mathrm{cv}} = h_{\mathrm{cv}}\Delta T \tag{20}$$

where $h_{\mathrm{cv}}$ is the convective heat transfer coefficient of air between gaps or at the outermost surfaces. $\Delta T$ is the temperature difference. Churchill and Chu derived the following equation for natural convection adjacent to a vertical plane [25]:

$$h_{\mathrm{cv,in}} = \frac{Nu \cdot \kappa_{\mathrm{air}}}{H} \tag{21}$$

among which $Nu$, $\kappa_{\mathrm{air}}$, and $H$ represent the mean Nusselt number, the thermal conductivity of air, and the PAM sample's height. As for Nusselt number [26,27], it follows,

$$\mathrm{Nu} = 0.68 + 0.670(\mathrm{Ra}\Psi)^{1/4}; \qquad \mathrm{Ra} \leq 10^9 \tag{22}$$

$$\mathrm{Nu} = 0.68 + 0.670(\mathrm{Ra}\Psi)^{1/4}(1 + 1.6 \times 10^8 \mathrm{Ra}\Psi)^{1/12}; \quad 10^9 \leq \mathrm{Ra} \leq 10^{12} \tag{23}$$

where $\Psi = \left[1 + (0.492/Pr)^{9/16}\right]^{-16/9}$; Ra is Rayleigh number and Ra=GrPr, where Gr is Grashof number (Gr) and the Prandtl number (Pr) [26,27]. For the convective heat transfer in the refined air gap [28],

$$h_{\mathrm{cv,gap}} = \frac{t_{\mathrm{gap}}}{\mathrm{Nu} \cdot k_{\mathrm{air}}} - h_{\mathrm{cd,gap}} \tag{24}$$

Nusselt number [28] can be calculated by,

$$Nu = \left[0.0605\mathrm{Ra}^{1/3}, \left(1 + \left[\frac{0.104\mathrm{Ra}^{0.293}}{1 + (6310/\mathrm{Ra})^{1.36}}\right]^3\right)^{1/3}, 0.243\left(\frac{\mathrm{Ra}}{H/t}\right)^{0.272}\right]_{\max} \tag{25}$$

in which, *H/t* is the gap aspect ratio (height to width). The thermal resistance can be calculated as

$$R = \frac{\Delta T}{q}. \tag{26}$$

Finally, the *U*-factor can be calculated as

$$U = \left[R_{\mathrm{ex}} + R_{\mathrm{cd,G}} + \sum_{i=1}^{N}\left(R_{\mathrm{r,gap,i}}{}^{-1} + R_{\mathrm{cd,gap,i}}{}^{-1} + R_{\mathrm{cv,gap,i}}{}^{-1}\right)^{-1} + N \cdot R_{\mathrm{cd,P}} + \left(R_{\mathrm{r,in}}{}^{-1} + R_{\mathrm{cv,in}}{}^{-1}\right)^{-1}\right]^{-1} \tag{27}$$

where the exterior heat transfer resistance $\boldsymbol{R}_{\mathbf{ex}}$ is set to 1/30 $m^2 \cdot K/W$ according to ASTM C1199.

## 4 Results and discussion

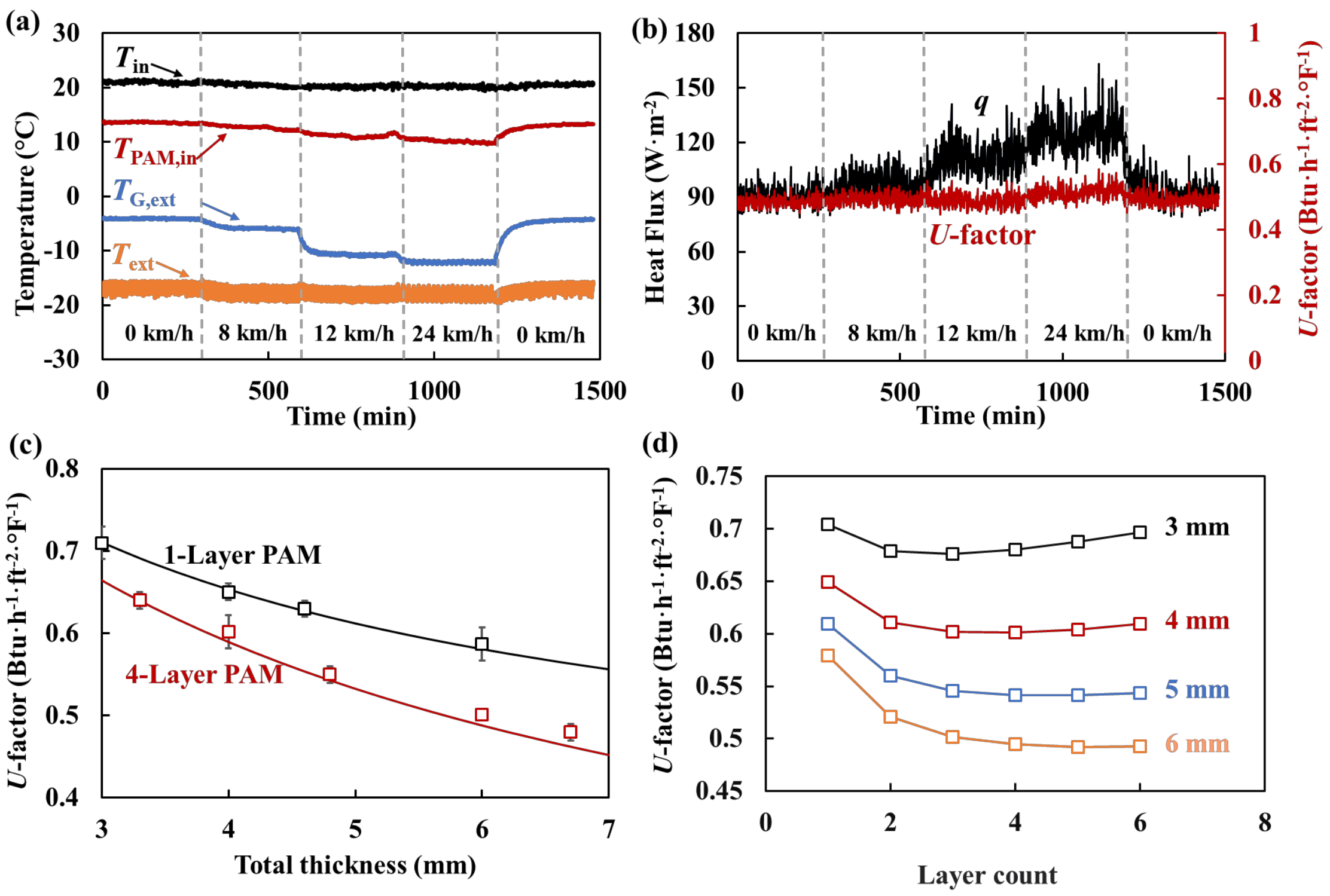

**Figure 4** The effects of the exterior air velocity on the temperature difference and the *U*-factor. (a) $T_{\mathrm{ex}}$, $T_{\mathrm{in}}$, $T_{\mathrm{PAM,ex}}$, and $T_{\mathrm{PAM,in}}$ and (b) the heat flux ($q$) and the steady-state *U*-factor ($t_{\mathrm{PAM}}$=6 ±0.1 mm). The wind speed is specified with the unit of km/h. (c) The *U*-factor as a function of the total thickness of 1-layer and 4-layer PAM. The structure of the 1-layer PAM is somewhat similar to Fig.2(a), except that there is only 1 PET film separated from the glass pane by an air gap [15]. (d) The calculated relationship between the *U*-factor and the layer count. The overall PAM thickness ranges from 3 mm to 6 mm, indicated by the number next to each curve.

Figure 4(a) and (b) show the measurement result of $T_{\mathrm{ex}}$, $T_{\mathrm{in}}$, $T_{\mathrm{G,ex}}$, $T_{\mathrm{PAM,in}}$, and the heat flux ($q$) under different wind speeds, for a 4-layer PAM sample. Its total thickness is 6 mm. When the wind speed was increased, the temperature difference and the heat flux across the PAM sample became larger, while the *U*-factor did not vary much, because the *U*-factor is calculated based on the standard heat transfer coefficient on the exterior surface ($h_{ST,ex}$= 30 W/(m$^2$·K)) per ASTM C1199 and NFRC-102. The *U*-factor was calculated by Eq. (2), around 0.5 Btu/(h· ft$^2$ · °F).

Figure 4(c) shows the *U*-factor of a 4-layer and a 1-layer PAM sample as a function of the overall PAM thickness. The solid lines are the numerical predication, and the data dots are the measurement results. Clearly, the *U*-factor is reduced when the total sample thickness is increased. The multilayer structure has better thermal insulation properties than the single layer structure. However, for a given total PAM thickness, when we further increase the layer count to larger than 4, the total thickness of air would be reduced, which resulted in an increase in the *U*-factor, as shown in Figure 4(d). For PAM design with the thickness around 3~6 mm, the optimum layer count is 4.

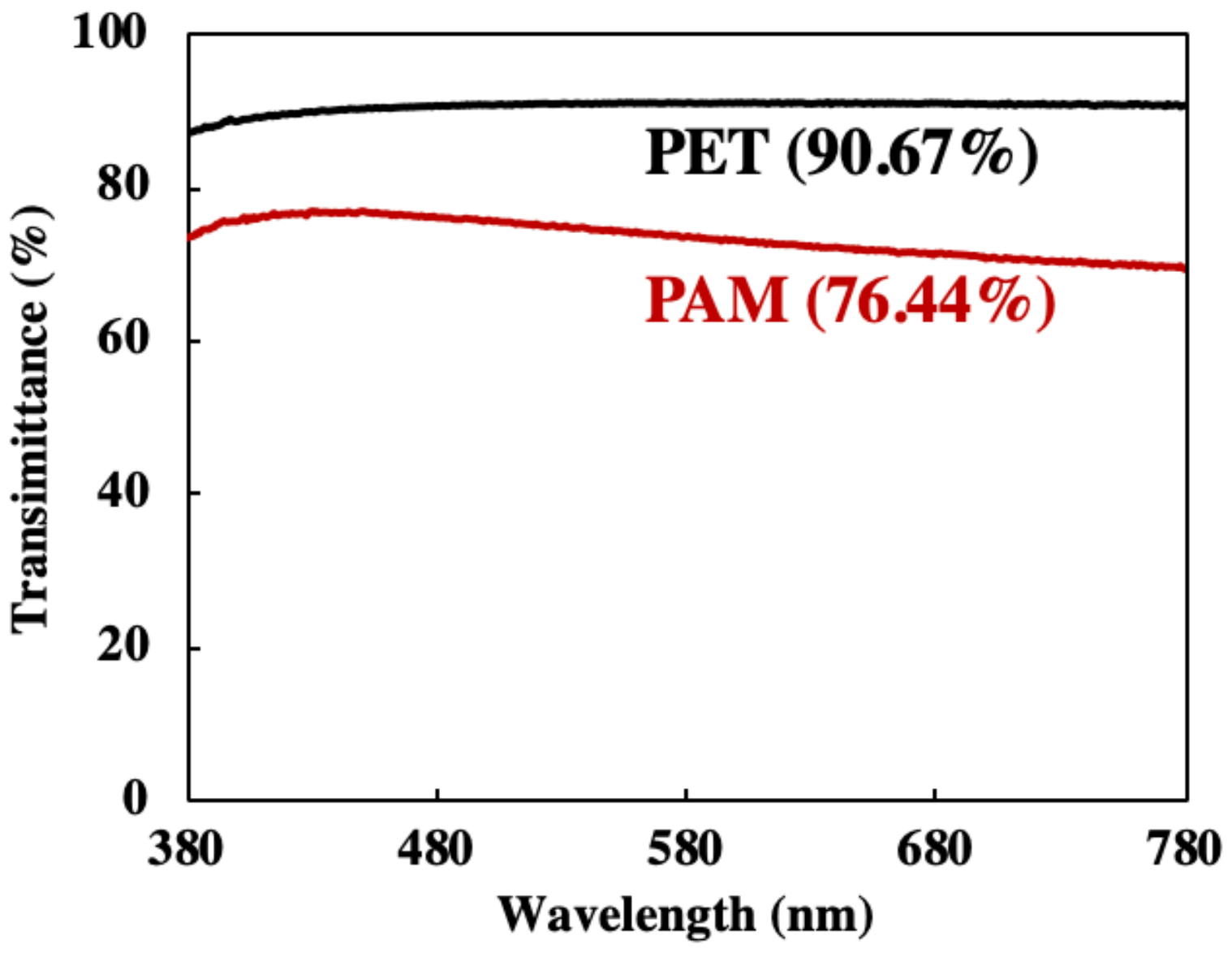

**Figure 5** The visual transmittance of a pristine PET film and a 4-layer PAM sample. The numbers in the brackets give the average $V_T$ over the spectrum range.

**Table 1** Summary of the optical properties

| Optical properties | | Pristine PET film | 4-layer PAM |
|---|---|---|---|
| $V_T$ (%) | | 90.7 | 76.44 |
| Haze (%) | | 0.2 | 1.6 |
| CRI | Color coordinates | (0.339319, 0.352434) | (0.334388, 0.347274) |
| | CRI value | 99.66 | 95.5 |

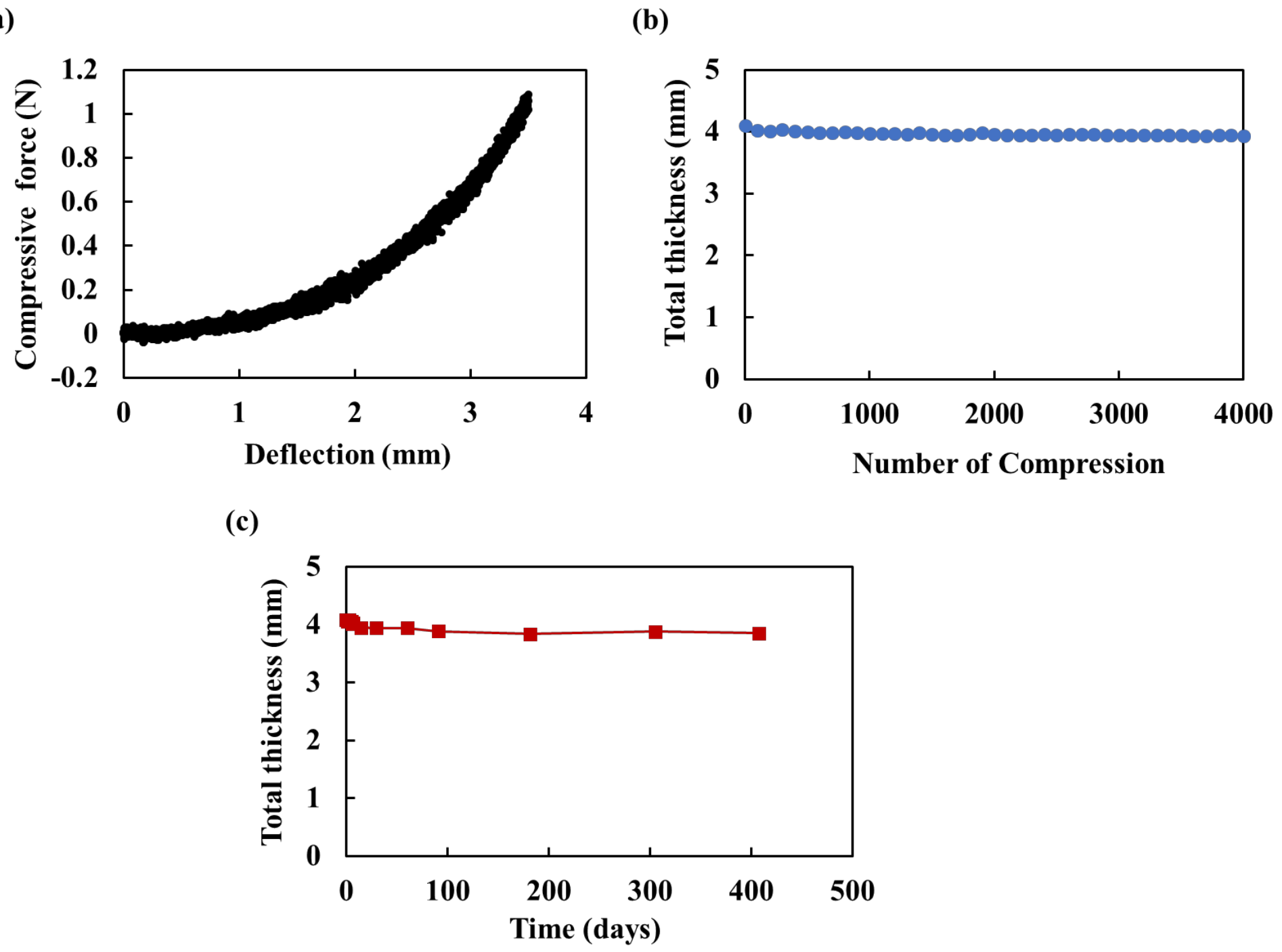

**Figure 6** (a) A typical compression curve of a 4-layer PAM. The stability of total air gap thickness upon (b) 4000 cycles of repeated flattening and (c) exposure for more than one year in lab air.

Figure 5 and Table 1 show the measured optical properties of a pristine PET film and a typical PAM sample. The visual transmittance ($V_T$) of a single PET film is 90.67%, and that of PAM is 76.44%, which is better than most of low-emissivity films [13]. Although $V_T$ of PAM is ~15% lower than that of a single PET layer, the overall thermal insulating properties is much improved. The haze of PAM is lower than 2%. Both of the color coordinates and CRI values suggest that PAM has an adequate achromatic and neutral color sensation, offering a desirable illumination with achromatic sensation.

Figure 6(a) shows a typical compression curve of a 0.3×0.3 m 4-layer PAM sample, with the initial thickness ($t_{PAM}$) being ~4 mm. The force required to flatten the PAM sample is around 1.1 N. After 4000 times of repeated flattening, the hollow multilayer structure can still be restored instantaneously when the load is removed, as indicated in Figure 6(b). The satisfactory structural robustness may be related to the large aspect ratio, the surface electrification, as well as the in-plane tensile stress. When the air gaps are compressed, the maximum strain in the polymer films is less than 0.3%, well within the linear elastic limit of PET [29]. Figure 6(c) indicates that the air gaps are stable up to over 1 year in lab air.

We also investigated the on-site assembly and installation procedure of PAM. The suggested steps include: sectioning of polymer films to the size of window pane, film clamping and stretching, spacer insertion, adhesive curing, trimming the edges, and mounting onto the window pane. In general, the like-charged PET films would not adhere to each other. They will be shear-cut into size and shape. The clamp is a portable device formed by metal bars, which holds and separates the polymer films. The clamped PAM set is affixed on a stretcher, and the polymer films are subject to a slight tension force. A small in-plane tensile stress is sufficient to render the films planar. Next, adhesive-sprayed spacer bars are inserted into the film stack along the edges. The structure is secured, as the adhesive is cured at room temperature for about 10 min. Finally, the edges of the PAM are trimmed, and the entire set is attached to a window pane, by using adhesives on the outer spacer bars. No adhesives are needed across the inner layer surface and the glass pane surface.

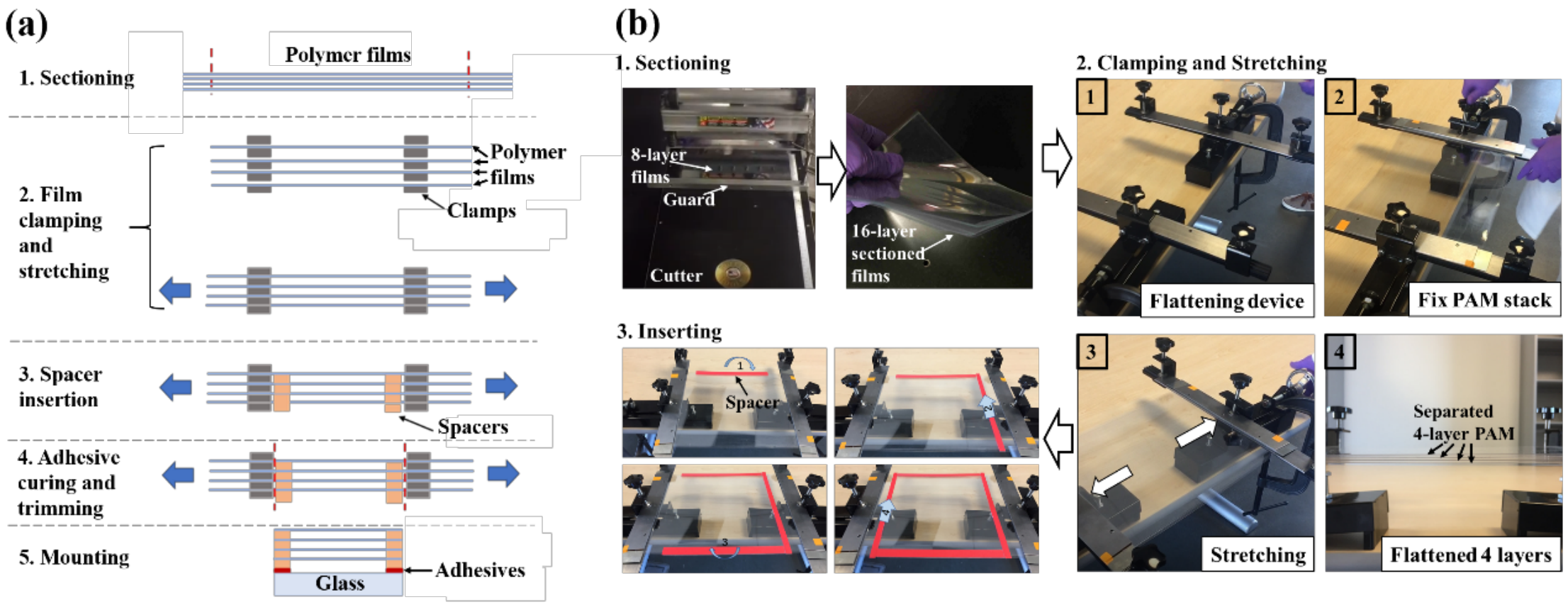

**Figure 7** (a) Schematic of the suggested installation process of a 4-layer PAM. (b) Photos of film sectioning, film clamping and stretching, and spacer insertion.

## 5. Concluding remarks

To summarize, in order to enhance the building window energy efficiency, we investigated the transparent, low-haze, and thermally insulating polymer-air multilayer (PAM) structure. Four flat 125-μm-thick PET films were separated by air gaps. The film separation was achieved by a set of spacer bars at the edges, and stabilized by slight stretching and surface electrification. The air gaps were 0.5~1 mm thick. The 4-layer PAM could have a low thermal transmittance ($U$-factor) less than 0.5 Btu/(h·ft$^2$·°F), as the conductive, convective, and radiative heat transfer was suppressed. The visual transmittance and haze were higher than 70% and lower than 2%, respectively. The PAM samples were flexible and resilient, stable upon more than 4,000 cycles of flattening and exposure to lab air for more than 1 year. This technology not only is relevant to retrofitting single-pane windows, but also may be useful for enhancing IGUs, soundproofness, etc.

**Acknowledgement:** This study was supported by the Advanced Research Projects Agency – Energy (ARPA-E) under Grant No. DE-AR0000737.

## References

[1] US Department of Energy, Research and Development Roadmap for Emerging HVAC Technologies, US Dep. Energy, Build. Technol. Off. Washington, DC. (2014) 121.

[2] U.S. DoE, Buildings energy databook, Energy Effic. Renew. Energy Dep. (2011).

[3] Single -Pan Highly Insulating Efficient Lucid Designs (SHIELD) Program Overview, (2014) 1–13.

[4] T. Gao, B.P. Jelle, T. Ihara, A. Gustavsen, Insulating glazing units with silica aerogel granules: The impact of particle size, Appl. Energy. 128 (2014) 27–34.

[5] U. Berardi, The development of a monolithic aerogel glazed window for an energy retrofitting project, Appl. Energy. 154 (2015) 603–615.

[6] Y.-L. He, T. Xie, Advances of thermal conductivity models of nanoscale silica aerogel insulation material, Appl. Therm. Eng. 81 (2015) 28–50.

[7] H. Maleki, L. Durães, A. Portugal, An overview on silica aerogels synthesis and different mechanical reinforcing strategies, J. Non. Cryst. Solids. 385 (2014) 55–74.

[8] R.P. Patel, N.S. Purohit, A.M. Suthar, An overview of silica aerogels, Int. J. ChemTech Res. 1 (2009) 1052–1057.

[9] M. Reim, W. Körner, J. Manara, S. Korder, M. Arduini-Schuster, H.-P. Ebert, J. Fricke, Silica aerogel granulate material for thermal insulation and daylighting, Sol. Energy. 79 (2005) 131–139.

[10] R. Baetens, B.P. Jelle, A. Gustavsen, Aerogel insulation for building applications: a state-of-the-art review, Energy Build. 43 (2011) 761–769.

[11] X.-D. Wang, D. Sun, Y.-Y. Duan, Z.-J. Hu, Radiative characteristics of opacifier-loaded silica aerogel composites, J. Non. Cryst. Solids. 375 (2013) 31–39.

[12] Y. Zhong, R. Kou, M. Wang, Y. Qiao, Synthesis of centimeter-scale monolithic SiC nanofoams and pore size effect on mechanical properties, J. Eur. Ceram. Soc. 39 (2019) 2566–2573.

[13] C. Schaefer, G. Bräuer, J. Szczyrbowski, Low emissivity coatings on architectural glass, Surf. Coatings Technol. 93 (1997) 37–45.

[14] R.J. Martın-Palma, L. Vazquez, J.M. Martınez-Duart, Silver-based low-emissivity coatings for architectural windows: Optical and structural properties, Sol. Energy Mater. Sol. Cells. 53 (1998) 55–66.

[15] R. Kou, Y. Zhong, J. Kim, Q. Wang, M. Wang, R. Chen, Y. Qiao, Elevating low-emissivity film for lower thermal transmittance, Energy Build. 193 (2019) 69–77.

[16] N. Sándor, J. Schanda, Visual colour rendering based on colour difference evaluations, Light. Res. Technol. 38 (2006) 225–239.

[17] How Much Does Window Replacement Cost? | Angie's List, (n.d.). https://www.angieslist.com/articles/how-much-does-window-replacement-cost.htm (accessed March 22, 2019).

[18] Y. Zhong, R. Kou, M. Wang, Y. Qiao, Electrification Mechanism of Corona Charged Organic Electrets, J. Phys. D. Appl. Phys. (2019).

[19] R. Kou, Y. Zhong, Y. Qiao, Effects of anion size on flow electrification of polycarbonate and polyethylene terephthalate, Appl. Phys. Lett. 115 (2019) 73704.

[20] W. Elenbaas, Heat dissipation of parallel plates by free convection, Physica. 9 (1942) 1–28.

[21] T. Smith, J. Guild, The CIE colorimetric standards and their use, Trans. Opt. Soc. 33 (1931) 73.

[22] C. Cie, Commission internationale de l'eclairage proceedings, 1931, Cambridge Univ. Cambridge. (1932).

[23] J. Schanda, The concept of colour rendering revisited, in: Conf. Colour Graph. Imaging, Vis., Society for Imaging Science and Technology, 2002: pp. 37–41.

[24] T.L. Bergman, F.P. Incropera, D.P. DeWitt, A.S. Lavine, Fundamentals of heat and mass transfer, John Wiley & Sons, 2011.

[25] A. Handbook, Fundamentals, Am. Soc. Heating, Refrig. Air Cond. Eng. Atlanta. 111 (2001).

[26] F.M. White, I. Corfield, Viscous fluid flow, McGraw-Hill New York, 2006.

[27] J.L. Wright, A correlation to quantify convective heat transfer between vertical window glazings, ASHRAE Trans. 102 (1996).

[28] A. Bejan, Convection heat transfer, John wiley & sons, 2013.

[29] K.K. Chawla, M.A. Meyers, Mechanical behavior of materials, Prentice Hall Upper Saddle River, 1999.